\documentclass[superscriptaddress,
twocolumn,showpacs,preprintnumbers,amsmath,amssymb]{revtex4}
\usepackage{graphicx}
\usepackage{dcolumn}
\usepackage{bm}
\usepackage{latexsym}
\begin{document}
\title{Quality control by a mobile molecular workshop: quality versus quantity} 
\author{Ajeet K. Sharma}
\author{Debashish Chowdhury{\footnote{Corresponding author(E-mail: debch@iitk.ac.in)}}}
\affiliation{Department of Physics, Indian Institute of Technology,
Kanpur 208016, India.}
\begin{abstract}
Ribosome is a molecular machine that moves on a mRNA track while, 
simultaneously, polymerizing a protein using the mRNA also as the  
corresponding template. We define, and analytically calculate, 
two different measures of the efficiency of this machine. However, 
we arugue that its performance is evaluated better in terms of the 
translational fidelity and the speed with which it polymerizes a 
protein. We define both these quantities and calculate these 
analytically. Fidelity is a measure of the quality of the products 
while the total quantity of products synthesized in a given interval 
depends on the speed of polymerization. We show that for synthesizing 
a large quantity of proteins, it is not necessary to sacrifice the 
quality. We also explore the effects of the quality control 
mechanism on the strength of mechano-chemical coupling. We suggest 
experiments for testing some of the ideas presented here. 
\end{abstract}
\maketitle
\section{Introduction} 

For cyclic machines with {\it finite cycle time}, the {\it efficiency 
at maximum power output} \cite{curzon75,broeck05,seifert} is a suitable 
indicator of its performance \cite{linke05}. For molecular motors, 
the {\it Stokes efficiency} \cite{oster,astumian} is an alternative 
measure of performance. However, not all machines are designed for the 
sole purpose of performing mechanical work. In this paper we consider a 
class of machines whose main function is to synthesize a hetero-polymer, 
subunit by subunit, using another hetero-polymer as the corresponding 
template. 

The conceptual framework that we develop here is generally applicable 
to all machines which drive template-dictated polymerization. 
But, for the sake of concreteness, we formulate the theory here for a 
specific machine, namely, the ribosome \cite{frank}; it polymerizes 
a protein using a messenger RNA (mRNA)  as the corresponding template 
and the process is referred to as {\it translation} (of genetic message).  
The subunits of a mRNA are nucleotides, whereas those of a protein are 
amino acids. The correct sequence of amino acids  to be selected by the 
ribosome is dictated by the corresponding sequence of the codons 
(triplets of nucleotides). The ribosome also uses the  mRNA template 
as the track for its movement. In each step, the ribosome moves 
forward by one codon on its track, and the protein gets elongated by 
one amino acid. Ribosome has a quality control system to minimize 
translational error by rejecting incorrect incoming amino acid subunits.

For a ribosome, the average speed $V$ of polymerization of a protein 
and the fidelity $\phi$ of translation, rather than efficiency and 
power output, are the primary indicators of its performance. We define 
these quantities, as well as a few others, quantitatively and calculate 
these analytically. Contrary to naive expectations, we show that a 
higher rate of elongation of the protein does not necessarily sacrifice 
the translational fidelity.

\section{The model} 

A ribosome consists of two interconnected parts called the large and 
the small subunits. The small subunit binds with the mRNA track and 
decodes the genetic message of the codon, whereas the polymerization 
of the protein takes place in the large subunit. The operations of the 
two subunits are coordinated by a class of adapter molecules, called 
transfer RNA (tRNA). One end of a tRNA helps in the decoding process 
by matching its anti-codon with the codon on the mRNA, while its other 
end carries an amino acid subunit; in this form the complex is called 
an amino-acyl tRNA (aa-tRNA).

\begin{figure}[t]
\begin{center}
\includegraphics[angle=-90,width=0.85\columnwidth]{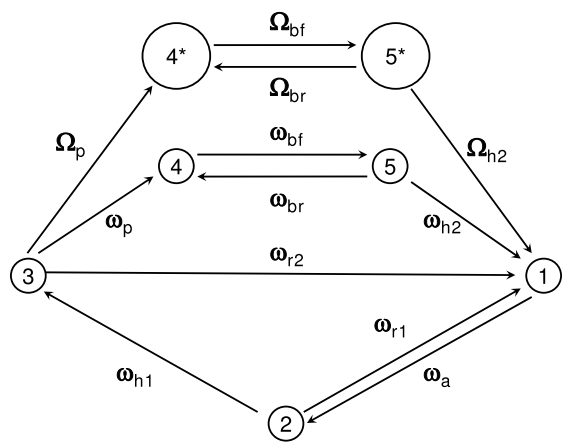}
\end{center}
\caption{Pictorial depiction of the full chemo-mechanical cycle 
of a single ribosome (see the text for details). 
}
\label{fig-ribomodel}
\end{figure}

The three main steps of each cycle of a ribosome 
are as follows: (i) selection of the cognate  aa-tRNA, (ii) formation 
of the peptide bond between the amino acid brought in by the selected 
aa-tRNA and the elongating protein, and (iii) translocation of the 
ribosome by one codon. However, some of these steps consist of 
important sub-steps. Moreover, the branching of paths from a well 
defined mechano-chemical state can give rise to the possibility of 
more than one cyclic pathway for a ribosome in a given cycle. 
Neither of these features was captured by the earlier models 
\cite{basu07,ciandrini10}.
The complete set of states and the pathways for a single ribosome in 
{\it our model} are shown in Fig.\ref{fig-ribomodel}. This model may be 
regarded as an extension of the Fisher-Kolomeisky generic model for 
molecular motors \cite{fishkolo}.

The selection of aa-tRNA actually consists of two sub-steps. In 
the first sub-step, among the tRNAs, which arrive at the rate 
$\omega_{a}$, non-cognate ones are rejected, at the rate 
$\omega_{r1}$, because of codon-anticodon mismatch. The second 
sub-step implements a kinetic proofreading mechanism for screening 
out the near-cognate tRNAs; this sub-step is irreversible and 
involves hydrolysis of a Guanosine triphosphate (GTP) molecule (at the 
rate $\omega_{h1}$). Near-cognate tRNAs are rejected at the rate 
$\omega_{r2}$. But, if the tRNA is cognate, the protein gets elongated 
by the addition of the corresponding amino acid (at the rate 
$\omega_{p}$). However, occasionally a near-cognate tRNA escapes 
potential rejection and the ribosome erroneously incorporates it into 
the protein (at a rate $\Omega_{p}$). The subsequent translocation step 
actually consists of important sub-steps. The first sub-step is a {\it 
reversible} relative (Brownian) rotation of the two subunits with 
respect to each other (with rates $\omega_{bf}$ and $\omega_{br}$ 
along the correct path and with rates $\Omega_{bf}$ and $\Omega_{br}$ 
along the wrong path). In the state labeled by $4$ (and $4^{*}$), 
the tRNAs are in the so-called ``hybrid'' configuration \cite{frank}, 
the details of which are not required for our purpose here. The 
second sub-step, which is {\it irreversible}, is driven by the 
hydrolysis of a GTP molecule; this sub-step leads to the coordinated 
movement of the tRNAs within the ribosome and the forward stepping 
of the ribosome by one codon on its mRNA track (at the rates 
$\omega_{h2}$ and $\Omega_{h2}$ along the correct and wrong paths, 
respectively). Further detailed identification of the states labeled 
in Fig.\ref{fig-ribomodel} by the integers \cite{scdwell} is not 
needed for our purpose here. 
 
Thus, following the selection of a cognate aa-tRNA, the correct pathway 
is $1 \to 2 \to 3 \to 4 \to 5 \to 1$. In contrast, if a non-cognate 
aa-tRNA is picked up, the most probable pathway is $1 \to 2 \to 1$. 
However, if the aa-tRNA is near cognate, then the pathway could be 
either $1 \to 2 \to 3 \to 1$ (successful kinetic proofreading)  
or $1 \to 2 \to 3 \to 4^{*} \to 5^{*} \to 1$ (incorporation of a 
wrong amino acid).

In principle, the model does not necessarily require any relation 
among the rate constants. However, throughout this paper we 
assume that 
\begin{equation}
\omega_{r2} + \Omega_{p} = C, 
\label{eq-constraint} 
\end{equation}
a constant, because the more stringent the proofreading is, the 
fewer will be the wrong proteins produced.

We define the fraction  
\begin{equation}
\phi= \dfrac{\omega_p}{\omega_p+\Omega_p}
\end{equation}
as a measure of translational {\it fidelity}. The {\it error ratio}     
$\epsilon =\Omega_p/(\omega_p+\Omega_p) = 1 - \phi$, 
is the fraction of wrong amino acids incorporated in a protein. 
We also define the {\it rejection factor} as 
\begin{equation}
{\cal R} = \biggl(\dfrac{\omega_{r1}}{\omega_{r1}+\omega_{h1}}\biggr)\biggl(\dfrac{\omega_{r2}}{\omega_{r2}+\omega_{p}+\Omega_{p}}\biggr). 
\end{equation} 
Note that the average velocity $V$ of a ribosome is also identical 
to the average rate of elongation of the protein that it polymerizes. 
A higher $V$ results in a larger ``quantity'' of the protein after a 
given interval. On the other hand, the parameter $\phi$ characterizes 
the ``quality'' of the final product while ${\cal R}$ characterizes 
the ``effort'' of the quality control system in screening out the 
non-cognate tRNA (including near-cognate tRNA).

\section{Results} 

Suppose that ${\cal P}_{\mu}(t)$ is the probability of finding the ribosome 
in the ``chemical'' state $\mu$ at time $t$, irrespective of its 
position on the mRNA track. The obvious normalization condition is 
$\sum_{\mu=1}^{5} {\cal P}_{\mu}+{\cal P}_4^*+{\cal P}_5^*=1$.
Solving the master equations for ${\cal P}_{\mu}(t)$ in the steady-state,   
we get the average velocity 
\begin{equation}
V= {\ell}_c (\omega_{h2} {\cal P}_5 + \Omega_{h2} {\cal P}_5^*) 
= {\ell}_c K_eff\biggl(1+\dfrac{\Omega_p}{\omega_p}\biggl)
\label{eq-v}
\end{equation} 
where ${\ell}_c$ is the length of a codon and
\begin{eqnarray}
\dfrac{1}{K_{eff}} &=& \dfrac{1}{\omega_{a}}\biggl(1+\frac{\omega_{r1}}{\omega_{h1}}\biggr)\biggl(1+\frac{\omega_{r2}}{\omega_{p}}\biggr) +\dfrac{1}{\omega_{h1}}\biggl(1+\frac{\omega_{r2}}{\omega_{p}}\biggr) \nonumber \\ 
&+&\dfrac{1}{\omega_{p}} + \dfrac{1}{\omega_{bf}}\biggl(1+\frac{\omega_{br}}{\omega_{h2}}\biggr) + \dfrac{1}{\omega_{h2}}\nonumber\\
&+&\biggl(\dfrac{\Omega_p}{\omega_p}\biggr)\biggr[\dfrac{1}{\omega_{a}}\biggl(1+\frac{\omega_{r1}}{\omega_{h1}}
\biggr)+\dfrac{1}{\omega_{h1}} \nonumber\\
&&~~~~~~~~+\dfrac{1}{\Omega_{bf}}\biggl(1+\frac{\Omega_{br}}{\Omega_{h2}}\biggr)+\dfrac{1}{\Omega_{h2}}\biggr]
\end{eqnarray}
Note that in the special case $\Omega_{p}=0$, 
different terms of $K_{eff}^{-1}$ are the average time spent by 
the ribosome in different mechano-chemical states.

\begin{figure}[t]
\begin{center} 
(a)\\
\includegraphics[angle=-90,width=0.75\columnwidth]{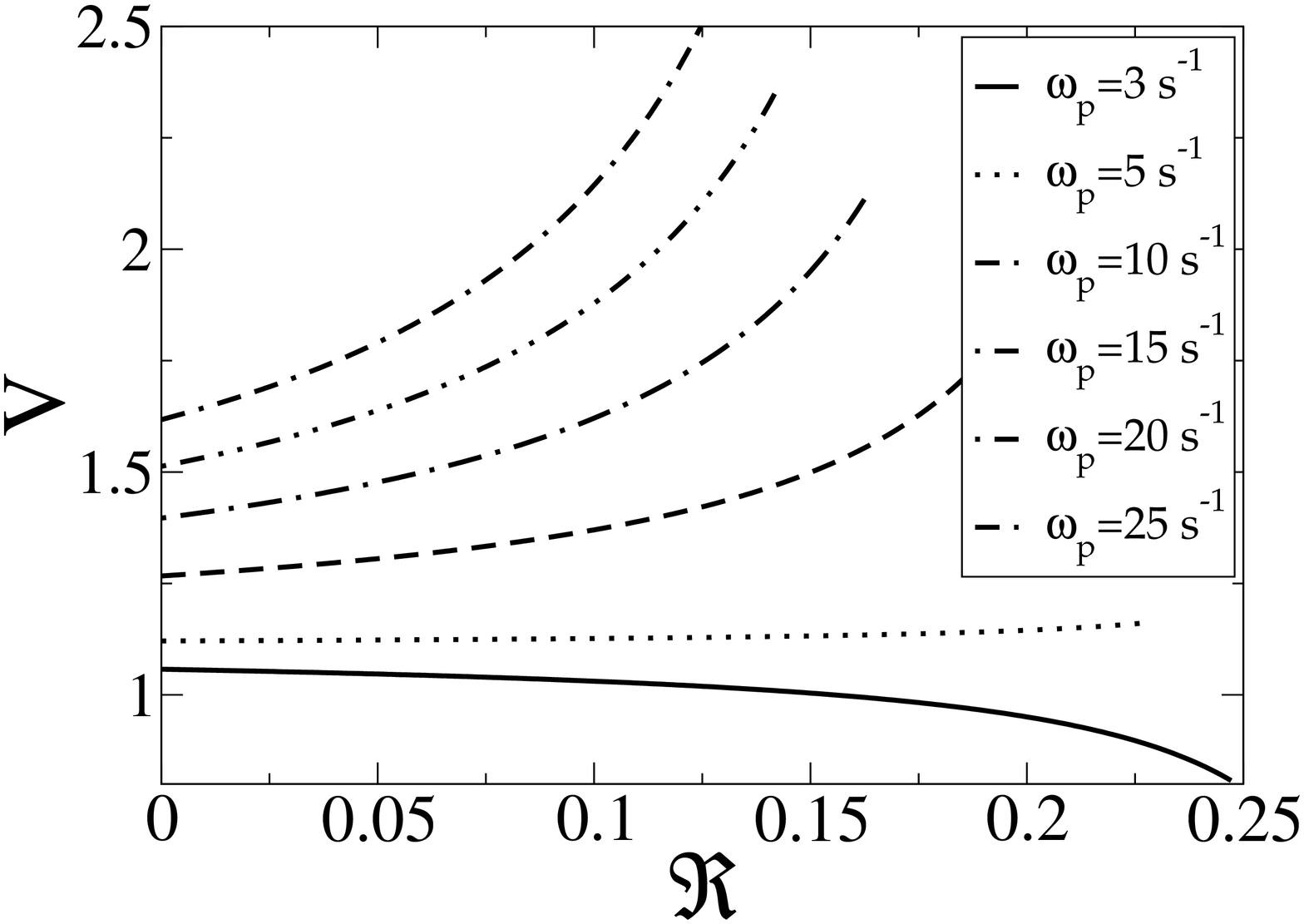}\\
(b)\\
\includegraphics[angle=-90,width=0.75\columnwidth]{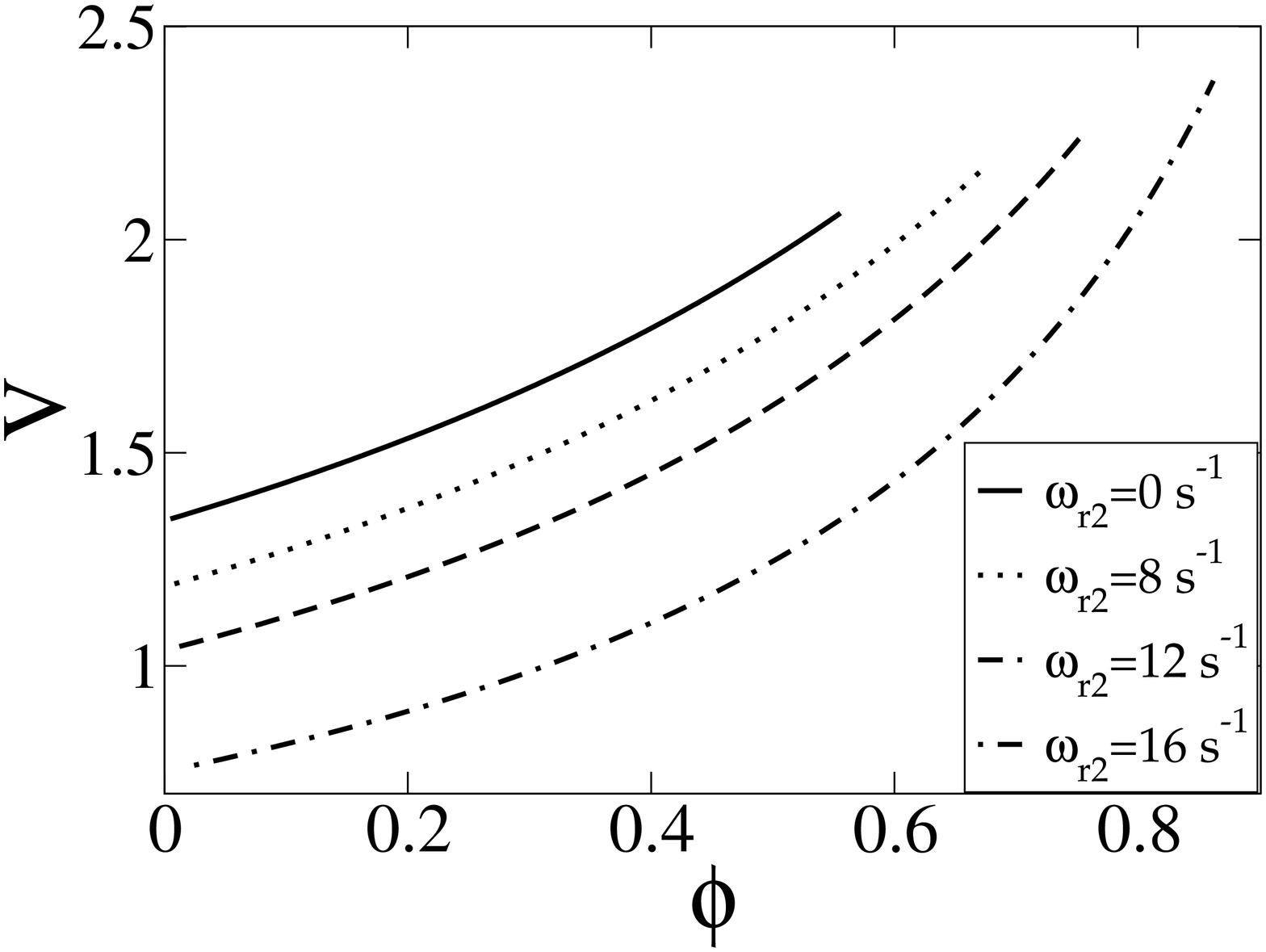}\\
(c)\\ 
\includegraphics[angle=-90,width=0.75\columnwidth]{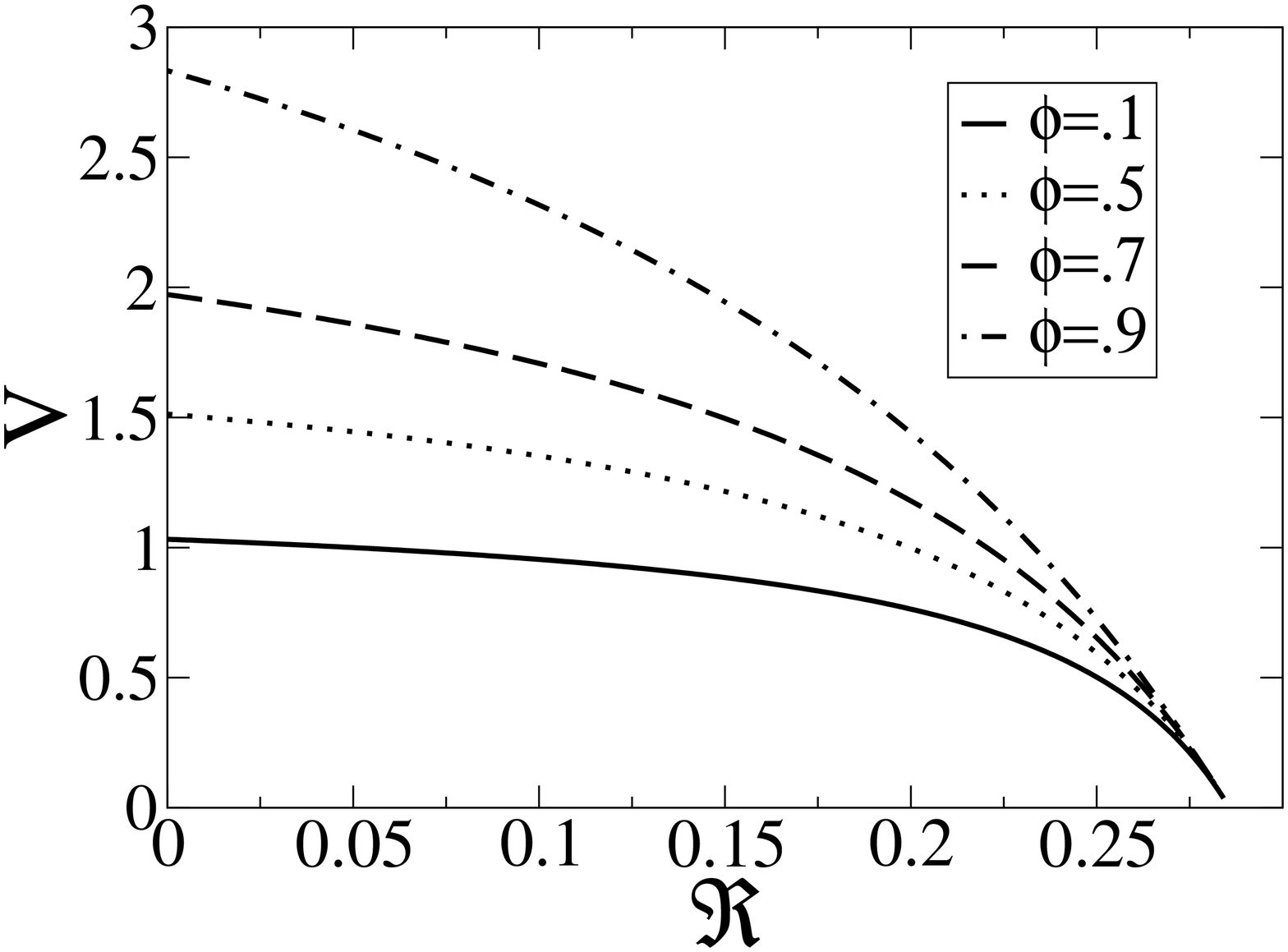}\\
\end{center}
\caption{The average velocity $V$ of a ribosome (in the units of 
``codons per second''), i.e., average rate of elongation of a 
protein (in the units of ``amino acids per second''), is plotted 
against (a) the rejection factor ${\cal R}$, (for six different 
fixed values of the parameter $\omega_{p}$), (b) the fidelity 
$\phi$ (for four different fixed values of the parameter 
$\omega_{r2}$), and (c) the rejection factor ${\cal R}$ (for four 
different fixed values of the fidelity $\phi$). 
}
\label{fig-vvsrphi}
\end{figure}

\begin{figure}[t]
\begin{center} 
\includegraphics[angle=-90,width=0.75\columnwidth]{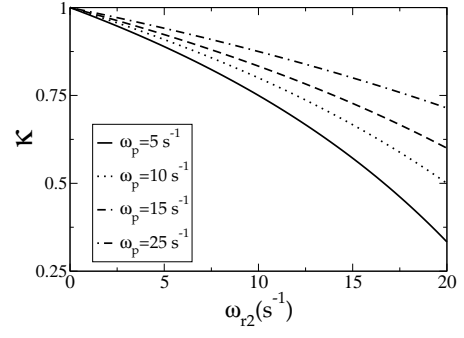}\\
\end{center}
\caption{The mechano-chemical coupling strength $\kappa$ is 
plotted against $\omega_{r2}$ for four different fixed values 
of the parameter $\omega_{p}$.
}
\label{fig-kvswr}
\end{figure}

\begin{figure}[t]
\begin{center}
\includegraphics[width=0.9\columnwidth]{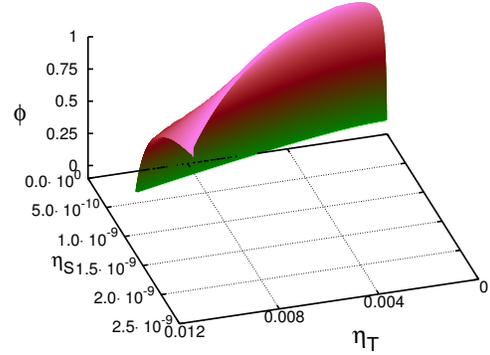}
\end{center}
\caption{(Color online) Thermodynamic efficiency $\eta_T$ and Stokes efficiency 
$\eta_S$ are plotted against the fidelity $\phi$.}
\label{fig-petes}
\end{figure}

The direct transition $3 \to 1$ gives rise to 
``slippage'' \cite{hillbook}. Therefore, the coupling between 
the chemical input and the mechanical output is ``loose'' \cite{oosawa00}. 
The ratio of the mechanical flux to chemical flux \cite{note} can be 
taken as a measure of the strength of the mechano-chemical coupling 
$\kappa$: 
\begin{equation}
\kappa = \frac{2(P_{5} \omega_{h2} + P_{5}^{*} \Omega_{h2})}{P_{2} \omega_{h1}+P_{5} \omega_{h2} + P_{5}^{*} \Omega_{h2}}
\label{eq-kappa}
\end{equation}
so that $\kappa = 1$ in the limit $\omega_{r2} = 0$. Similar quantitative 
measures of mechano-chemical coupling have been introduced earlier 
\cite{imre} in the general context of motors which can execute ``futile'' 
cycles of hydrolysis of nucleotide tri-phosphates (including GTP).
In the steady state of our model, 
\begin{equation}
\kappa = \frac{2 (\omega_{p} + \Omega_{p})}{2(\omega_{p} + \Omega_{p}) + \omega_{r2}} = \frac{2 (\omega_{p} + C - \omega_{r2})}{2(\omega_{p} + C) - \omega_{r2}}
\end{equation}

The input power consumed in proofreading is
\begin{eqnarray}
P_{in}^p = \biggr[\biggl(\dfrac{\omega_{r2}}{\omega_{r2}+\omega_p+\Omega_p}\biggr)\omega_{h1}P_2\biggr]\Delta\mu 
\end{eqnarray}
whereas the input power consumed in protein synthesis
\begin{eqnarray}
P_{in}^s =
\biggl[\biggl(\dfrac{\Omega_p+\omega_p}{\omega_{r2}+\Omega_p+\omega_p}\biggr)\omega_{h1}{\cal P}_2+\Omega_{h2}{\cal P}_5^*+\omega_{h2}{\cal P}_5\biggr]\Delta\mu
\end{eqnarray}
where $\Delta \mu$ is the free energy released by the hydrolysis of 
a single GTP molecule. Thus, the total power input is
\begin{eqnarray}
P_{in} =
\biggr(\dfrac{\omega_{r2}+2\omega_p+2\Omega_p}{\omega_p}\biggr) K_{eff}\Delta\mu
\end{eqnarray}

We define the thermodynamic efficiency $\eta_T$ and the Stokes 
efficiency $\eta_S$ by the relations 
\begin{eqnarray}
\eta_T = \frac{F V}{P_{in}}, ~{\rm and}~ \eta_S = \frac{\gamma V^{2}}{P_{in}}
\label{eq-defeta}
\end{eqnarray}
where $F$ is the load force and $\gamma~ V$ is the phenomenological 
form of the viscous drag on the ribosome. One of the limitations 
\cite{linke05} of definitions (\ref{eq-defeta}) is that it 
implies vanishing thermodynamic efficiency as the external load 
force vanishes although, in reality, the motor continues to work 
against the viscous drag. $\eta_{S}$ is the appropriate measure 
of the efficiency of a molecular motor when $F$ vanishes. From 
definitions (\ref{eq-defeta}) $\eta_S=(\gamma V/F) \eta_T$, and 
using expression (\ref{eq-v}) for $V$, we get
\begin{equation}
\eta_T=\dfrac{F {\ell}_c(\Omega_p+\omega_p)}{(\omega_{r2}+2\omega_p+2\Omega_p) \Delta \mu}
\end{equation}
Moreover, we assume the standard form \cite{fishkolo} of $F$-dependence 
of $\omega_{h2}(F)$ and $\Omega_{h2}(F)$: the values of these rate 
constants in the absence of $F$ are multiplied by the factor 
$exp(-\delta~F {\ell}_c/k_B T)$ where $0 < \delta < 1$. 
For plotting the graphs, we have used the parameter 
values $F = 1$pN, $\Delta \mu = 10 k_B T$, ${\ell}_c=0.9$nm, $\delta=0.5$, 
$\omega_a =25.0$s$^{-1}$, $\omega_{h1}= 25.0$s$^{-1}$, 
$\omega_{h2}(0) = 25.0$s$^{-1}$, $\omega_{br} = \omega_{bf} = 10.0$s$^{-1}$, 
$\omega_p =25.0$s$^{-1}$, and $\gamma =60$pN.s.m$^{-1}$. 
The parameters $\omega_{r1}$ and $\omega_{r2}$ have been varied over 
a wide range always keeping $\omega_{r2} + \Omega_{p} = 20$s$^{-1}$.

The average rate $V$ of elongation of a protein is plotted 
against the rejection factor ${\cal R}$ and fidelity $\phi$ 
in Figs.\ref{fig-vvsrphi}(a) and (b), respectively (note the 
unit used for plotting $V$). The increase of 
$\omega_{r2}$ increases the rejection factor ${\cal R}$. Therefore, 
naively, one would expect $V$ to decrease with increasing ${\cal R}$. 
However, because of constraint (\ref{eq-constraint}), the increase 
in $\omega_{r2}$ is compensated by decreasing $\Omega_{p}$. Therefore, 
instead of decreasing, $V$ can increase with ${\cal R}$ provided 
$\omega_{p}$ is sufficiently large (see Fig.\ref{fig-vvsrphi}(a)).

Moreover, contrary to naive expectation, $V$ increases with 
increasing $\phi$ as long as $\omega_{r2}$ is kept fixed (see 
Fig.\ref{fig-vvsrphi}(b)). However, for a given $\phi$, $V$ 
decreases with increasing $\omega_{r2}$; this decrease is caused 
by the increasing frequency of the futile cycles in which a 
molecule of GTP is hydrolyzed in kinetic proofreading and the 
aa-tRNA is rejected.  
In order to emphasize the interplay of ${\cal R}$ and $\phi$ 
directly, in Fig.\ref{fig-vvsrphi}(c), we plot $V$ against 
${\cal R}$ for several different values of the parameter $\phi$. 
Clearly, for a given $\phi$, $V$ decreases monotonically with 
${\cal R}$. Similar monotonic decrease in $V$ with increasing 
$\omega_{r2}$ demonstrates that the ``slippage'' caused by the 
kinetic proofreading weakens the mechano-chemical coupling 
$\kappa$ (see Fig.\ref{fig-kvswr}). 

The variation of the efficiencies $\eta_T$ and $\eta_S$ with 
$\phi$ is shown in the three-dimensional plot of Fig.\ref{fig-petes}. 
Taking cross sections of this diagram parallel to the $\eta_T-\eta_S$ 
plane for several different constant values of $\phi$ (not shown 
in any figure) we find that $\eta_T$ increases monotonically 
with $\eta_S$, but the increase is sublinear. This is a 
consequence of the fact that although $\phi$ depends only on 
the ratio $\Omega_p/\omega_p$, $V$ depends separately also on 
$\omega_p$ through its dependence on $K_{eff}$.

\section{Summary and conclusion} 

Intracellular machines which carry out template-dictated polymerization 
of macromolecules can be regarded as molecular motors \cite{gelles98,frank}.
Obviously, as we have shown here, appropriate measures of efficiency 
can be defined and calculated for such motors. However, the performance 
of these machines is evaluated better in terms of {\it fidelity} and 
rate of polymerization, rather than efficiency and power output. Using 
our theoretical model for the kinetics of translation, we show that,  
contrary to the widespread belief, a higher rate of polymerization of 
a protein does not necessarily compromise the translational fidelity. 
We have also analyzed the interplay of the ``futile cycles'' arising 
from kinetic proofreading and the consequent ``looseness'' of the 
mechano-chemical coupling. 

For carrying out the calculations analytically, we ignored the possible 
variations of the rate constants from one codon to another. Therefore, 
our prediction can be tested {\it in-vitro} using an artificially 
synthesized sequence-homogeneous mRNA (i.e., mRNA strand whose codons 
are all identical, except for the start and stop codons). In the 
surrounding medium only two species of amino acid subunits should be 
available, one of these two species is cognate whereas the other is not. 
The rate constants $\omega_{r1}$, $\omega_{r2}$, $\Omega_{p}$, 
$\Omega_{bf}$, $\Omega_{br}$ and $\Omega_{h2}$ can be varied by replacing 
the non-cognate amino acid subunits with another distinct species which 
is also non-cognate.

The performance of all the intracellular machines of template-dictated 
polymerization can be characterized and evaluated within the general 
conceptual framework developed here. However, the results for polymerases 
\cite{schliwa}, which polymerize DNA and RNA, would differ from those 
of ribosomes because of the differences between their respective 
mechano-chemical cycles.

\noindent We thank Joachim Frank for constructive criticism of an 
earlier draft of the manuscript.



\begin{thebibliography}{99}
\bibitem{note} Since $\omega_{h2}$ (and, similarly, $\Omega_{h2}$) 
accounts for a transition that is both ``chemical'' (hydrolysis of a 
GTP molecule) and ``mechanical'' (forward translocation) in nature, 
the corresponding flux appears both in the numerator and denominator 
of Eq.(\ref{eq-kappa}) defining $\kappa$.
\end{thebibliography}
\end{document}